\documentclass[11pt]{article}
\usepackage{amsmath,amsthm,latexsym,amssymb,amsfonts,amsthm}
\usepackage{graphicx, color, calc, url}
\usepackage[list=true, font=large, labelfont=bf,
labelformat=brace, position=top]{subcaption}

\makeatletter \@addtoreset{equation}{section} \makeatother

\title{Entanglement entropy of physical states in hypercuboidally truncated spin foam quantum gravity}

\author{Benjamin Bahr$^1$\\
\small $^1$ II Institute for Theoretical Physics\\
\small University of Hamburg\\
\small  Luruper Chaussee 149\\
\small 22761 Hamburg, Germany
 }
\date{}

\begin{document}

\maketitle

\begin{abstract}
In this article we consider physical states in the hypercuboidal truncation of the EPRL-FK spin foam quantum gravity model. In particular, these states are defined on graphs which allow considering the entanglement entropy (EE) associated to the bipartition of space. We compute the EE numerically for some examples, and find that it depends on the coupling constants within the theory. We also find that there appears a maximum of the EE within the region of the coupling constant containing the non-Gaussian fixed point of the RG flow of the truncated model. We discuss the relation of this behaviour with the restoration of diffeomorphism symmetry at the fixed point. 
\end{abstract}

\section{Introduction}

Spin Foam models (SFM) are certain proposals for the construction of transition amplitudes for states defined on graphs. A prime example are the spin foam models for quantum gravity, which have been developed as expressions for the physical inner product of spin network states in loop quantum gravity (LQG), but also topological BF theory, or even (pure) lattice gauge theory, can be formulated in terms of spin foam models. \cite{Ashtekar:1994mh}\cite{Ashtekar:1995zh}\cite{Rovelli:1995ac}\cite{Reisenberger:1996pu}\cite{Baez:1999sr}\cite{Bahr:2010bs}\cite{Bahr:2012qj}

SFM therefore deliver proposals for physical states in LQG, in that they can be used to define a rigging map, i.e.~a bona fide projector from kinematical to physical states satisfying the constraints. \cite{Henneaux:1992ig}\cite{Giulini:1998kf} \footnote{As the SFM on the market in quantum gravity are usually derived directly from a quantization of the path integral for GR, and not from the canonical quantum theory, the relation between the two is still subject to discussion.}

As these SFM are defined on discrete structures, the question of the continuum limit naturally arises. This limit is captured in a refinement of both the bulk lattices (2-complex), as well as the boundary graphs, and leads to a notion of cylindrical consistency, allowing to construct the full continuum Hilbert space as an inductive limit over graphs \cite{Manrique:2005nn}\cite{Bahr:2009qc}\cite{Rovelli:2010qx}\cite{Bahr:2011aa}\cite{Bahr:2014qza}\cite{Dittrich:2014ala}\cite{Banburski:2014cwa}\cite{Dittrich:2016tys}. This programme is a form of background-independent renormalisation, in which the coarseness of lattices plays the role of the scale, since in quantum gravity usual parameters such as e.g.~lattice lengths are part of the dynamical fields themselves, which encode the geometry of space-time. In this framework there are several choices of renormalisation scheme, and the precise choice of boundary states, in particular ones stable under coarse graining, is an active field of research \cite{Delcamp:2016dqo}\cite{Delcamp:2016yix}\cite{Livine:2017xww}\cite{Livine:2019cvi}.

One of the most widely used models for the transition of LQG spin network states is the EPRL-FK model, which is defined on 2-complexes dual to 4d triangulations, and its KKL-extension to general 2-complexes, allowing the use of arbitrary polytopes \cite{Engle:2007wy}\cite{Freidel:2007py}\cite{Kaminski:2009fm}. It relies on a specific implementation of the so-called simnplicity constraints on topological $SO(4)$-BF theory, building on a classical equivalence of  GR with BF theory, in which the bivector field $B$ is constrained to be simple. This model has received much attention since its inception, although the question of its renormalisation is still very much open. 

In \cite{Bahr:2015gxa}, a specific truncation of the EPRL-FK-KKL model was introduced in order to construct a toy model, which serves as a laboratory for renormalisation\footnote{As a side product, using the hypercuboidal geometries in this truncation, it was realised that in the EPRL-FK model the volume-part of the simplicity constraints are insufficiently implemented, leading to non-metric degrees of freedom in the path integral. These have been understood to be linked to conformal matching of boundary faces of 3d polytopes in }. The model restricts the fluctuating geometries to specific (hyper-)cuboidal geometries \cite{Livine:2007vk}\cite{Bianchi:2010gc}. Interestingly, it was found that the RG flow already of this simple model is non-trivial, and induces a flow of the face amplitude, which governs the powers of volume factors in the path integral measure \cite{Bahr:2016hwc}\cite{Bahr:2017klw}. Using frustal geometries, it was found that the fixed point is non-Gaussian (NGFP), in the sense that it lies at specific non-zero values of Newton's coupling and the cosmological constant.

The NGFP separates two regions of phase space with vastly different geometric behaviour. Specifically, there are different (geometrically equivalent) states which receive different weights in terms of regular / irregular subdivision of polytopes. It is at the fixed point where these geometries are all treated equally, indicating a restoration of diffeomorphism symmetry. That this symmetry is broken in the EPRL-FK model has been known for some time \cite{Bahr:2009ku}. In particular, it is broken in Regge Calculus (RC), which arises in a certain limit of the EPRL-FK model, and while the symmetry is restored in classical RC even for flat configurations, it is broken even for those in the quantum theory due to the form of the path integral measure \cite{Bahr:2015gxa}. It was conjectured for some time that symmetries broken due to discretisation get restored at the coarse graining fixed point), and the properties of the NFGP are an indication for this mechanism in the 4d quantum gravity theory \cite{Bahr:2009qc}\cite{Bahr:2011uj}.

Still, many features of the NGFP are yet to be understood.\footnote{For instance, the question of a phase transition is still open.} To alleviate this somewhat, in this article we consider the entanglement entropy (EE) of physical states, at and away from the fixed point. EE is a very general concept, which is of great interest for general many-body systems \cite{Eisert:2008ur}. For a physical system with local degrees of freedom, is measures the entanglement of degrees of freedom inside of a spatial region $A$ with the ones outside of $A$. Here in particular the scaling behaviour of the EE is of interest: while generic states in the Hilbert space of a theory scale with the region volume, many ground states for interesting physical Hamiltonian operators scale with only the surface \cite{Srednicki:1993im}\cite{Eisert:2008ur}. It is this property which is used to identify and construct such states, for instance by a multiscale-entanglement renormalisation ansatz (MERA), or further developments building on this concept \cite{White:1992zz}\cite{Vidal:2007hda}. 

Also in LQG the concept of entropy has been considered in relation to black holes \cite{Ashtekar:1997yu}, and in terms of EE of bipartitie systems for quite some time \cite{Livine:2005mw}\cite{Livine:2006xk} \cite{Donnelly:2008vx}\cite{Delcamp:2016eya}\cite{Donnelly:2016auv}\cite{Livine:2017fgq}\cite{Feller:2017jqx}, in particular in view of isolated horizons. The spin network functions, which are defined on graphs thought of as embedded in (or building up) 3-dimensional space, have a geometrical interpretation which is ideally suited to discuss degrees of freedom associated to specific regions in space. However, for a single spin network, the only entropy between nodes is between Gauss-gauge degrees of freedom. If counting only gauge-invariant degrees of freedom, then the EE vanishes, and the state essentially factorises over the nodes of the graph (see section \ref{Sec:EE}).  

However, for physical states this picture changes. A physical state arises as the image of a kinematical one under the rigging map, and it can be represented as a superposition of different spin networks. The precise superposition depends on the parameters of the model, i.e.~on the coupling constants. 

In this article, we will compute the entanglement entropy $S_\text{EE}^{(\alpha)}$ for physical states in the hypercuboidal truncation of the EPRL-FK-KKL model. We will work in the large spin region of state space, in which the expressions for the path integral amplitude will become numerically manageable\footnote{The full amplitude is quite cumbersome to compute numerically, see e.g.~\cite{Dona:2019dkf}.}. We will then investigate $S_\text{EE}^{(\alpha)}$ near the fixed point, and discuss its behaviour in relation to the restoration of diffeomorphism symmetry.\\

The plan of the article is as follows: In section \ref{Sec:EPRL-Model} we recap the EPRL-FK-KKL model, as well as the hypercuboidal truncation. In section \ref{Sec:PIP} we discuss the construction of physical states, using a dynamical embedding as rigging map. In section \ref{Sec:EE}  we review the concept of entanglement entropy, and derive expressions for $S_\text{EE}^{(\alpha)}$ for different physical states, and in particular some scaling behaviour, which will help us to numerically compute the $\alpha$-dependence numerically in section \ref{Sec:Numerics}. Finally, we will interpret and discuss our findings in section \ref{Sec:Summary}.

\section{The EPRL-FK spin foam model}\label{Sec:EPRL-Model}

\begin{figure}
\begin{center}
\includegraphics[scale=0.75]{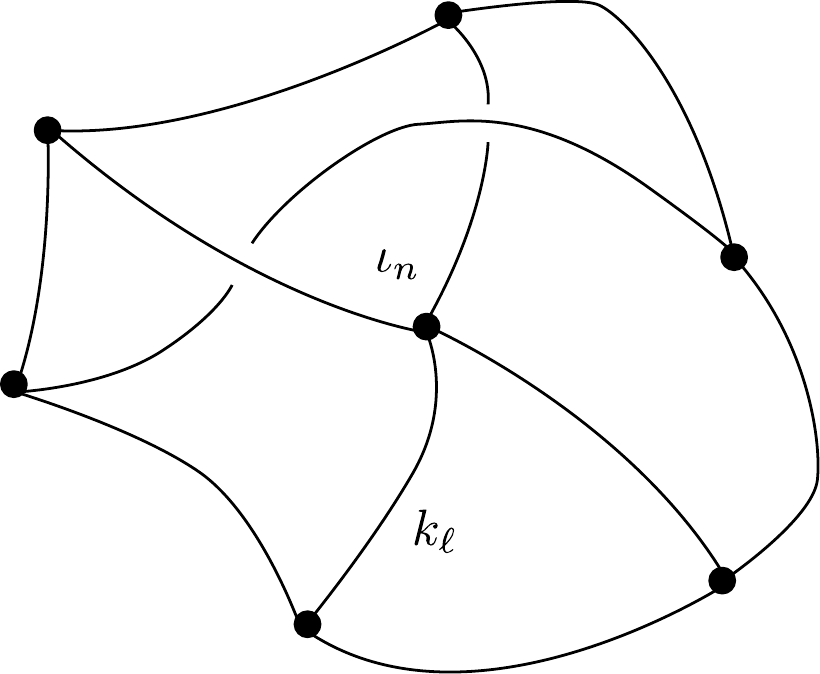}
\caption{Graphs $\Gamma$ with spins $k_\ell$ and invariant tensors $\iota_n$ among links and nodes consititute an orthonormals basis for the gauge-invariant Hilbert space of LQG.}\label{Fig:01}
\end{center}
\end{figure}

A spin network function on a graph $\Gamma$ is defined on an oriented graph $\Gamma$, and is labelled by a collection of spins $k_\ell\in\frac{1}{2}\mathbb{N}$ on the links $\ell$ of $\Gamma$, as well as invariant tensors
\begin{eqnarray}\label{Eq.Intertwiner_Space}
\iota_n\;\in\;\text{Inv}_{SU(2)}
\left[
\bigotimes_{[n,\ell]=1}V_{k_\ell}\otimes\bigotimes_{[n,\ell]=-1}V_{k_\ell}^\dag
\right]
\end{eqnarray}

\noindent along nodes $n$ in $\Gamma$, where $[n,\ell]=\pm 1$, depending on whether a link $\ell$ is outgoing / incoming to the node $n$. A widely-used overcomplete basis of the intertwiner spaces (\ref{Eq.Intertwiner_Space}) for fixed spins is given by the \emph{Livine-Speziale-coherent intertwiners}, which depend on $3d$ normal vectors $\vec{n}_\ell$ satisfying the \emph{closure constraint}

\begin{eqnarray}\label{Eq:Closure}
G_n\;=\;\sum_{[n,\ell]=1}k_\ell \vec{n}_{\ell}\;-\;\sum_{[n,\ell]=-1}k_\ell \vec{n}_{\ell}\;=\;0.
\end{eqnarray}

\begin{figure}
\begin{center}
\includegraphics[scale=0.5]{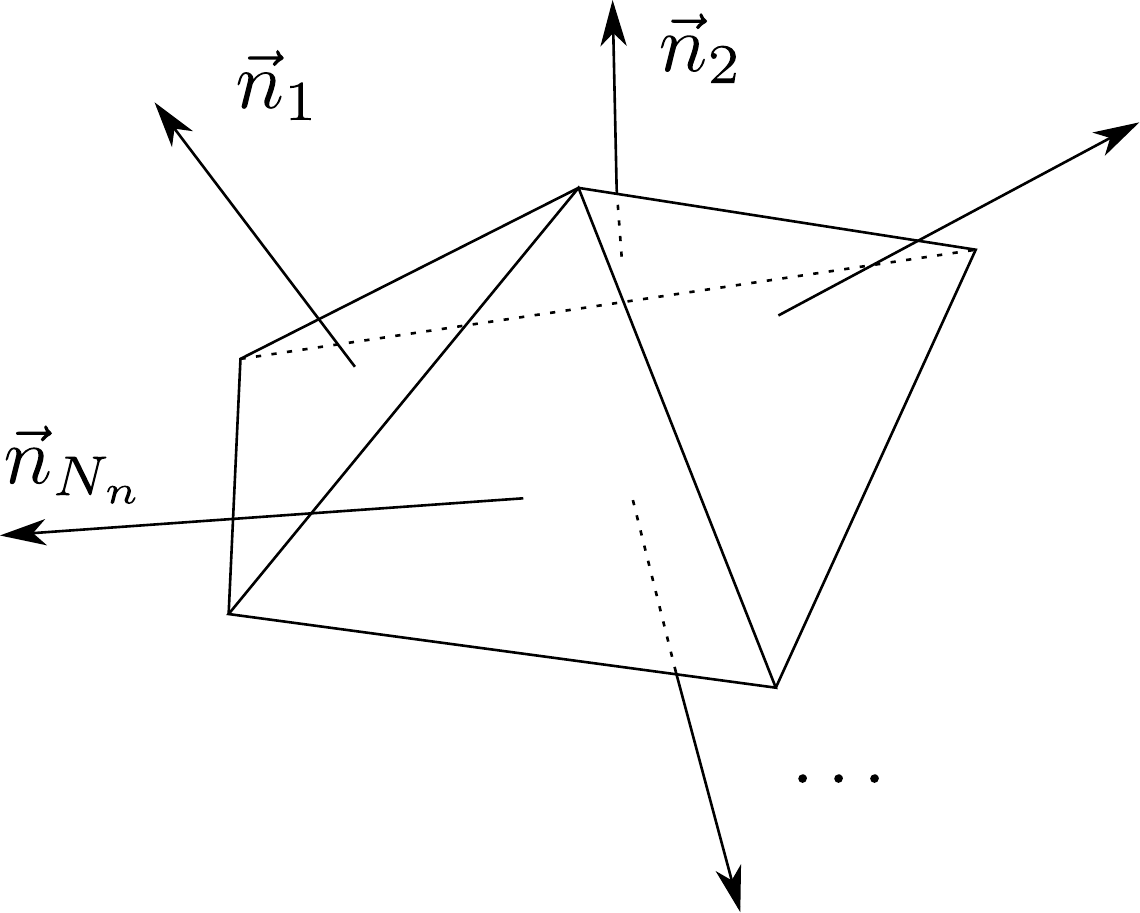}
\caption{Livinte-speziale-Intertwiners correspond to quantised $3d$ polytopes with fixed areas.}\label{Fig:02}
\end{center}
\end{figure}

\noindent With these, the Livine-Speziale coherent intertwiner is given by
\begin{eqnarray}\label{Eq:LS_Intertwiner}
\iota_\ell\;=\;\int_{SU(2)}dg\;g\triangleright
\left[
\bigotimes_{[n,\ell]=1}|k_\ell,\;\vec{n}_\ell\rangle\otimes\bigotimes_{[n,\ell]=-1}
\langle k_\ell,\;\vec{n}_\ell|
\right]
\end{eqnarray}

\noindent where $|k,\,\vec{n}\rangle=g_{\vec{n}}|k,\;k\rangle$ is the Perelomov coherent state defined by the action of $g_{\vec{n}}\in SU(2)$, the $SU(2)$-rotation which rotates $\vec{e}_z$ into $\vec{n}$, on the highest weight vector of the representation $k$.\footnote{If the normal vectors $\vec{n}_\ell$ do not satisfy the closure condition (\ref{Eq:Closure}), the state (\ref{Eq:LS_Intertwiner}) is still an intertwiner, although its norm is exponentially suppressed for large spins. In \ref{Conrady:2009px} it was shown that the resolution of identity can be restricted to those intertwiner satisfying (\ref{Eq:Closure}), if one includes an additional measure factor in the path integral. }

\begin{figure}
\begin{center}
\includegraphics[scale=0.45]{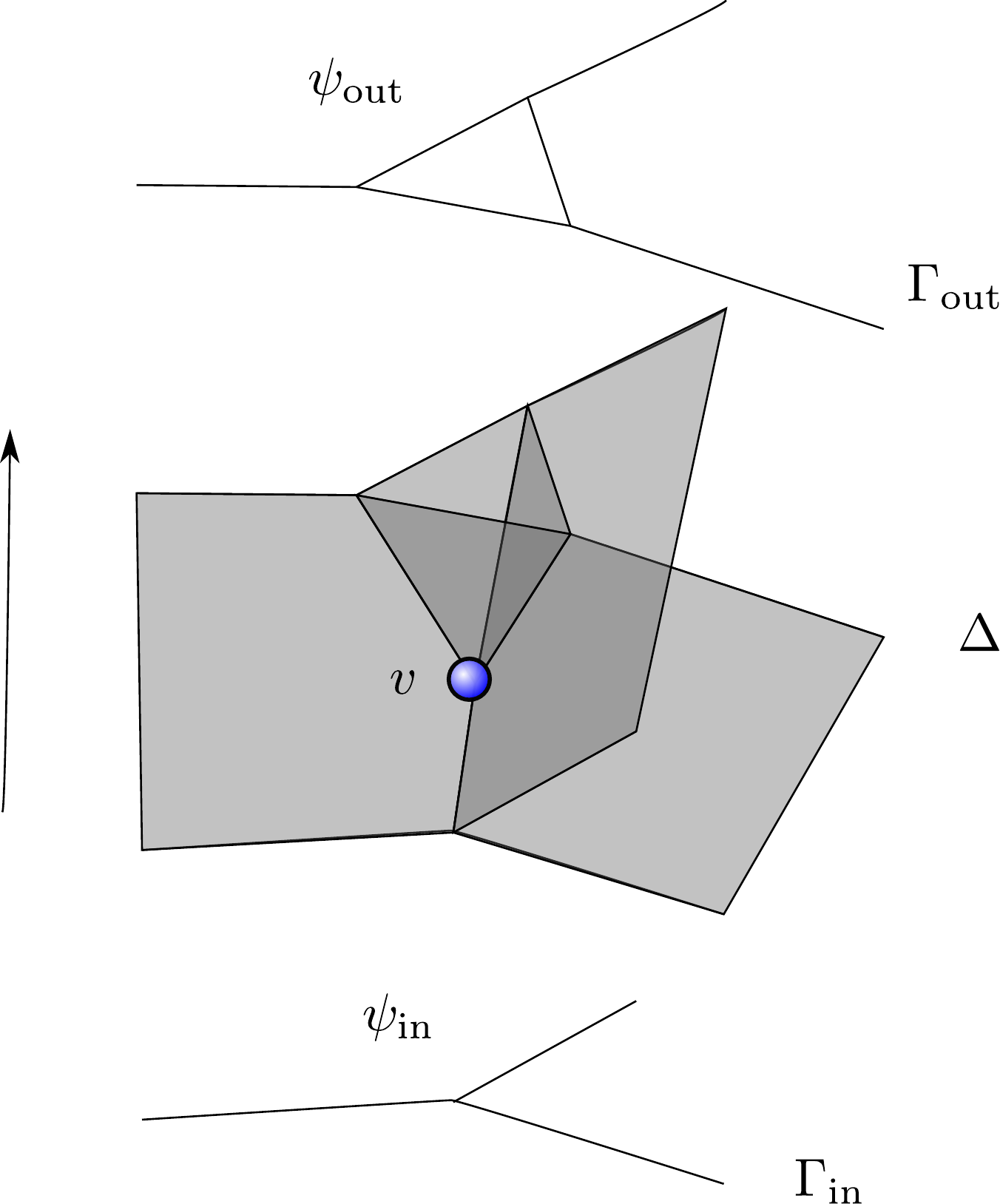}
\caption{Spin foam transition: a 2-complex $\Delta$ bounded by two graphs $\Gamma_\text{in}$ and $\Gamma_\text{out}$.}\label{Fig:03}
\end{center}
\end{figure}

The spin foam state sum is defined on a $2$-complex ${\Delta}$, consisting of 2d faces, 1d edges, and 0d vertices. The 2-complex functions as a cobordism between two graphs (see figure \ref{Fig:03}), which arise on its boundary, as those edges and vertices which touch only one face and edge, respectively. 

Consider an oriented 2-complex $\Delta$ with a (not necessarily connected) boundary graph $\Gamma$. Then a state is an assignment of spins $k_f$ to 2d faces $f$ of $\Delta$, and of intertwiners $\iota_e$ to edges $e$, from the tensor product of spins $k_f$ on faces $f$ meeting at $e$. The vertices and edges on the boundary form the nodes $n$ and links $\ell$ of $\Gamma$, and touch exactly one face $f_\ell$ and edge $e_n$ in the bulk respectively. Therefore, via $k_\ell\equiv k_{\ell_f}$ and $\iota_n\equiv\iota_{n_e}$ the state $\{k_f,\iota_e\}$ induces a spin network $\psi_{\Gamma,\{k_\ell\},\{\iota_n\}}$ on the boundary. The spin foam amplitude assigned to the state $\{k_f,\iota_e\}$ is given by
\begin{eqnarray}\label{Eq:SpinFoamStateSum}
\mathcal{Z}_\Delta[\psi_{\Gamma,\{k_\ell\},\{\iota_n\}}]
\;=\;
\sum_{j_f,\iota_e}\prod_{f\subset F_\text{bulk}}\mathcal{A}_f\prod_{e\in E_\text{bulk}}\mathcal{A}_e
\prod_{v\in V_\text{bulk}}\mathcal{A}_v\;\prod_{f_\ell\in F_\text{bdy}}\mathcal{B}_{f_\ell}\prod_{e_n\in E_\text{bdy}}
\mathcal{B}_{e_n},
\end{eqnarray}

\noindent where $\mathcal{A}_f$, $\mathcal{A}_e$ and $\mathcal{A}_v$ are, respectively, the face-, edge-, and vertex amplitude of the model, while the $\mathcal{B}_{f_\ell}$ and $\mathcal{B}_{e_n}$ are the boundary amplitudes, which are usually chosen in such a way that $\mathcal{Z}_\Delta$ behaves naturally under glueing \cite{Bahr:2010bs}.

The summation in (\ref{Eq:SpinFoamStateSum}) ranges only over spins and intertwiners in the bulk, while those on the boundary are being kept fixed, as they are determined by the boundary state. 

In this article we work with the EPRL-FK-KKL model, which amounts to a specific choice for the amplitudes, described in detail in \cite{Bahr:2015gxa}. The model depends on the Barbero-Immirzi parameter $\gamma$, which in our case we allow to take values in $\gamma\in(0,1)$. To be precise, the sum in (\ref{Eq:SpinFoamStateSum}) is restricted to range over those $k_f$ such that \footnote{This is a peculiarity of the Riemannian signature model The analogous condition for the Lorentzian version of the model does not restrict the allowed spins \cite{Engle:2007wy}\cite{Conrady:2010kc}. }
\begin{eqnarray}\label{Eq:SimplicitySpinCondition}
j_f^\pm\;:=\;\frac{|1\pm\gamma|}{2}k_f\;\in\;\frac{1}{2}\mathbb{N}.
\end{eqnarray}

\noindent In what follows, we work in the large spin regime where the sum over $k_f$ is approximated by integrals, and since the density of allowed $k_f$ within in $\frac{1}{2}\mathbb{N}$ is constant along its range, we can ignore this restriction of spins, since it just gives an overall density factor for $\mathcal{Z}_\Delta$.

In what follows we are only interested in the amplitudes for a specific subset of states, which comprise the hypercuboidal truncation of the model.

\subsection{Hypercuboidal truncation}

\begin{figure}
\begin{center}
\includegraphics[scale=0.45]{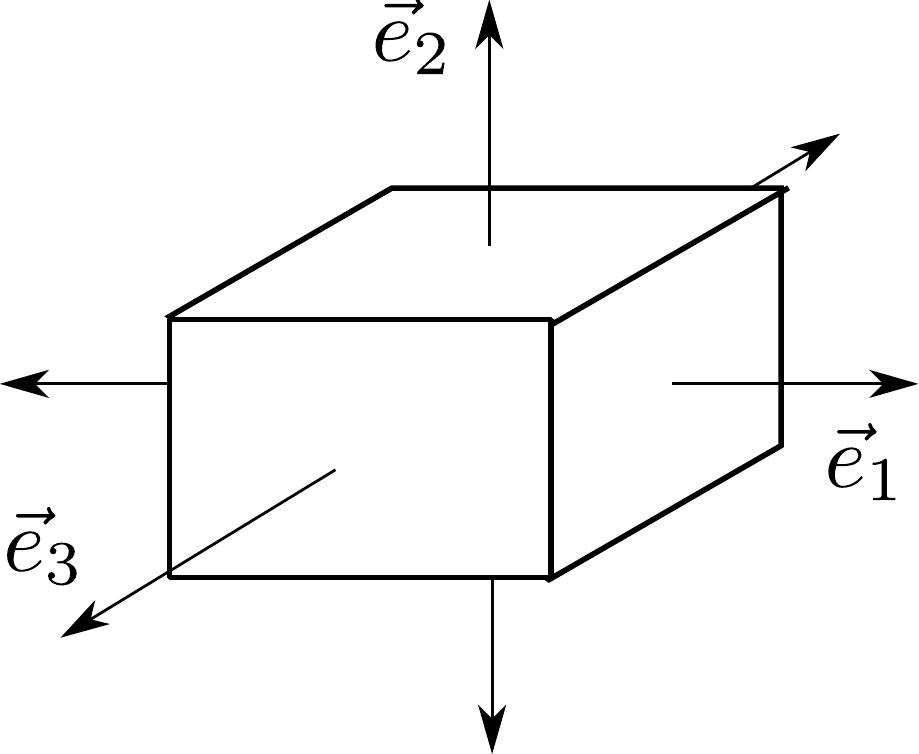}
\caption{Quantum cuboids as specific LS-intertwiners arising in the hypercuboidal truncation of the EPRL-FK-model.}\label{Fig:04}
\end{center}
\end{figure}

The model truncated on hypercuboids is essentially a restriction to a specific set of allowed spins and intertwiners on a 2-complex $\Delta$ dual to the 2-skeleton of a 4-dimensional hypercubic lattice. The intertwiners in question are so-called \emph{quantum quboids}, which in the large spin limit have the geometric interpretation of 3d cuboids with fixed areas (see figure \ref{Fig:04}). Each quantum cuboid is completely determined by three spins, and is given by
\begin{eqnarray}\label{Eq:QuantumCuboid}
\iota_{k_1,k_2,k_3}
\;:=\;
\int_{SU(2)}dg\triangleright\left[
\bigotimes_{i=1}^3|k_i,\,\vec{e}_i\rangle\otimes\langle k_i,\;\vec{e}_i|
\right]
\end{eqnarray}

\begin{figure}
\begin{center}
\includegraphics[scale=0.45]{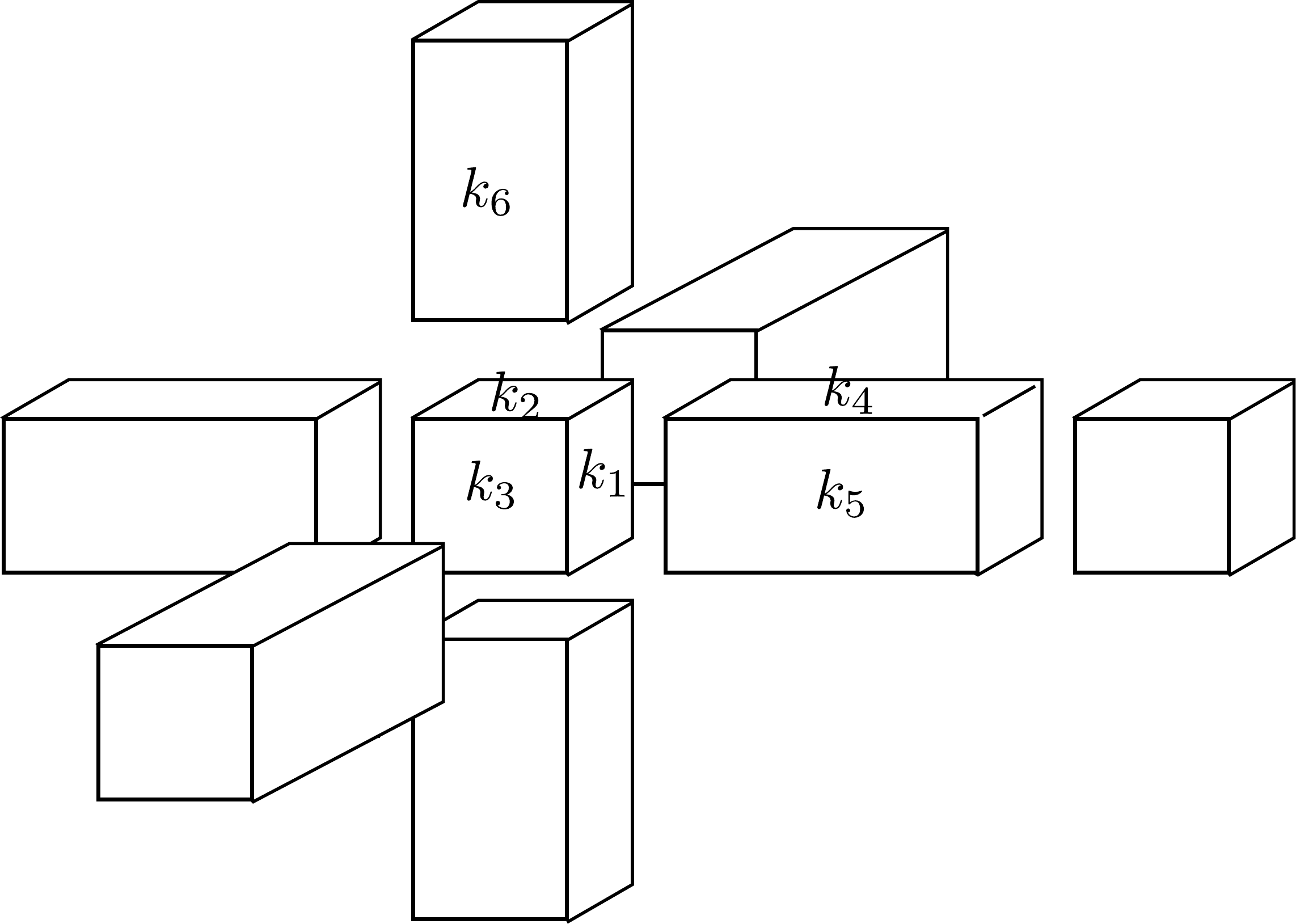}
\caption{Each hypercuboidal vertex is bounded by eight quantum cuboids. Thus a vertex amplitude depends on six spins $k_1,\ldots,k_6$.}\label{Fig:05}
\end{center}
\end{figure}

\noindent where $\vec{e}_i$ are the unit vectors pointing in the $i$-th direction in $\mathbb{R}^3$. The quantum cuboids are only defined when faces on opposite sides have equal spin, which restricts the sum in (\ref{Eq:SpinFoamStateSum}) to a highly symmetric set, which has been described in detail in \cite{Bahr:2015gxa}. Still, there are quasi-local propagating degrees of freedom, which have, however, no interpretation as curvature. At each vertex there are 24 faces meeting, but due to the high amount of symmetry, quadruples of them have equal spin. As a consequence, a vertex amplitude $\mathcal{A}_v$ depends only on six spins $k_1,\ldots,k_6$ (see figure \ref{Fig:05}). In the large spin limit, the asymptotic expression of the amplitude equals (up to a $k_i$-independent factor)
\begin{eqnarray}\label{Eq:VerteAmplitude}
\mathcal{A}_v(k_1,\ldots, k_6)
\;=\;
\left(\frac{1}{D}\,+\,\frac{1}{D^*}\right)^2,
\end{eqnarray} 

\noindent where $D$ depends on the six spins $k_i$ via
\begin{eqnarray*}
D^2\;&=&\;2 \big(k_1^2 (k_2 + k_4) + k_2 k_4 (k_2 + k_4) \\&+&
   k_1 (k_2^2 + (1 + i) k_2 k_4 + k_4^2)\big) \big(k_1^2 (k_3 + k_5) \\&+&
   k_3 k_5 (k_3 + k_5) + k_1 (k_3^2 + (1 + i) k_3 k_5 + k_5^2)\big) \big(k_3 k_4 k_5 \\&+&
   k_2 (k_4 k_5 + k_3 (k_4 + k_5))\big) \big(k_2^2 (k_3 + k_6) + k_3 k_6 (k_3 + k_6) \\&+&
   k_2 (k_3^2 + (1 + i) k_3 k_6 + k_6^2)\big) \big(k_4^2 (k_5 + k_6) \\&+&
   k_5 k_6 (k_5 + k_6) + k_4 (k_5^2 + (1 + i) k_5 k_6 + k_6^2)\big) \big(k_3 k_4 k_6 \\&+&
   k_1 (k_4 k_6 + k_3 (k_4 + k_6))\big) (k_2 k_5 k_6 + k_1 (k_5 k_6 + k_2 (k_5 + k_6))),
\end{eqnarray*}

\noindent where the branch cut to define $D$ is put on the negative real axis. The omitted prefactor in (\ref{Eq:VerteAmplitude}) contains all of the dependence of $\gamma$, and can be ignored in what follows. The edge amplitude is given by the inverse norm squared of the $SU(2)\times SU(2)$ intertwiner\footnote{This follows from the fact that we sum over normalised intertwiners, since the edge operator is a projector, see e.g.~\cite{Bahr:2015gxa} for details. } which can, up to an irrelevant factor, be written as
\begin{eqnarray}\label{Eq:EdgeAmplitude}
\mathcal{A}_e(k_1,k_2k_3)
\;=\;
(k_2+k_3)(k_3+k_1)(k_1+k_2).
\end{eqnarray}

\noindent The face amplitude is given by 
\begin{eqnarray}\label{Eq:FaceAmplitude}
\mathcal{A}_f(k)\;=\;k^{2\alpha},
\end{eqnarray}

\noindent where $\alpha$ is a free parameter in the model, which has been introduced in \cite{Bahr:2015gxa}. This parameter will play a crucial role in the following investigations. 

It should be noted that, in the large spin expressions, all explicit dependencies of the Barbero-Immirzi parameter $\gamma$ go into irrelevant prefactors, as all of the individual amplitude functions are homogeneous functions of the $j^\pm$ in (\ref{Eq:SimplicitySpinCondition}) of various degrees.

A finite 2-complex has eight boundaries (two in each major axis direction in $\mathbb{R}^4$), and a face $f$ touching only one of those gets assigned the boundary amplitude
\begin{eqnarray}
\mathcal{B}_f\;=\;\left(\mathcal{A}_f\right)^\frac{1}{2},
\end{eqnarray}

\noindent while a face $f$ touching two (i.e.~on a ``corner'' of the lattice) gets assigned
\begin{eqnarray}
\mathcal{B}_f\;=\;\left(\mathcal{A}_f\right)^\frac{1}{4},
\end{eqnarray}

\noindent since this is then only a quarter of a full rectangular face. The edges never end in corners, so we define for all boundary edges $e$ that
\begin{eqnarray}
\mathcal{B}_e\;=\;\left(\mathcal{A}_e\right)^\frac{1}{2},
\end{eqnarray}

\noindent which is just the inverse of the norm of the boosted quantum cuboid intertwiner at that edge. Due to the high amount of symmetry, one can split up and rearrange all the amplitudes, associating them to vertices, writing
\begin{eqnarray}
\mathcal{Z}\;=\;\sum_{k_f}\prod_v\hat{\mathcal{A}}_v
\end{eqnarray}

\noindent where the \emph{dressed vertex amplitude} is given by
\begin{eqnarray}
\hat{\mathcal{A}}_v\;:=\;\mathcal{A}_v\prod_{e\supset v}\left(\mathcal{A}_e\right)^\frac{1}{2}\prod_{f\supset v}\left(\mathcal{A}_f\right)^\frac{1}{4}.
\end{eqnarray}

\noindent This way, the boundary amplitudes are correctly taken care of.

\subsection{Geometricity of the vertex amplitude} 

The set of spins $k_f$ distributed among the faces of the lattice, which comply to the hypercuboidal symmetry, contains many elements with non-metric interpretation, in the sense that they do not allow for a reconstruction of the $4d$ metric from the spins. The presence of these configurations results from the insufficient implementation of the volume simplicity constraint on non-simplicial vertex amplitudes. These non-metric configurations also appear on more general vertices, and are generally characterised by the fact that, unlike twisted geometries \cite{Freidel:2010aq}\cite{Dittrich:2008va}\cite{Freidel:2013bfa}, they are not suppressed in the large-spin regime. They feature face-non-matching, but $2d$ angle-matching, i.e.~there is a conformal mismatch between touching faces, preventing glueing in $4d$ \cite{Bahr:2015gxa}\cite{Dona:2017dvf}\cite{Bahr:2017ajs}.

The conditions on the spins to remove these non-metric configurations can, for the hypercuboid, be formulated in terms of the Hopf link volume constraint in \cite{Bahr:2017ajs}, which result, for each vertex $v$ in the conditions
\begin{eqnarray}\label{Eq:GeometricityHypercuboidSpins}
k_1k_6\;=\;k_2k_5\;=\;k_3k_4.
\end{eqnarray}

\noindent For large values of $\alpha$ the non-metric configurations appear to be dynamically suppressed \cite{Bahr:2015gxa}, but in general one can demand their absence from the start. In what follows, we will consider both cases of present and absent non-metric degrees of freedom in the path integral. 

\subsection{Kinematical and dynamical embedding maps}

A central part of background-independent renormalization is the relation of degrees of freedom on different graphs. This is connected to the ``rescaling'' of degrees of freedom in the traditional context, and to the ``block spin transformations'' in the lattice theory context.

In SFM this is encoded in the embedding map, which maps the Hilbert space of a coarse graph $\Gamma$ to a refined graph $\Gamma'$. 
\begin{eqnarray}\label{Eq:EmbeddingMaps}
\phi_{\Gamma'\Gamma}\;:\;\mathcal{H}_{\Gamma}\;\longrightarrow\;\mathcal{H}_{\Gamma'},
\end{eqnarray}

\noindent where $\Gamma$ can be either the boundary of $\Delta$, or the boundary of a single vertex $v$ in $\Delta$, in order to renormalise single amplitudes. 

In \cite{Bahr:2016hwc}\cite{Bahr:2017klw}\cite{Bahr:2018gwf}, a kinematical embedding map was used which commutes with the electric fluxes of the EPRL-FK model. For a quantum cuboid state on the coarse graph $\Gamma$ with spins $K_I$ on squares $I$, and an embedding into quantum cuboid states with fine spins $k_i$, the map can be written as
\begin{eqnarray}\nonumber
\phi_{\Gamma'\Gamma}\psi_{\vec K}\;=\;\frac{1}{N_{\vec K}}\sum_{k_i}\left(\prod_{\text{coarse squares }I}\delta\big(K_I-\sum_{i\subset I}k_i\big)\right)\;\psi_{\vec k},\\[5pt]\label{Eq:EmbeddingMapQuboid}
\end{eqnarray}

\begin{figure}
\begin{center}
\includegraphics[scale=0.45]{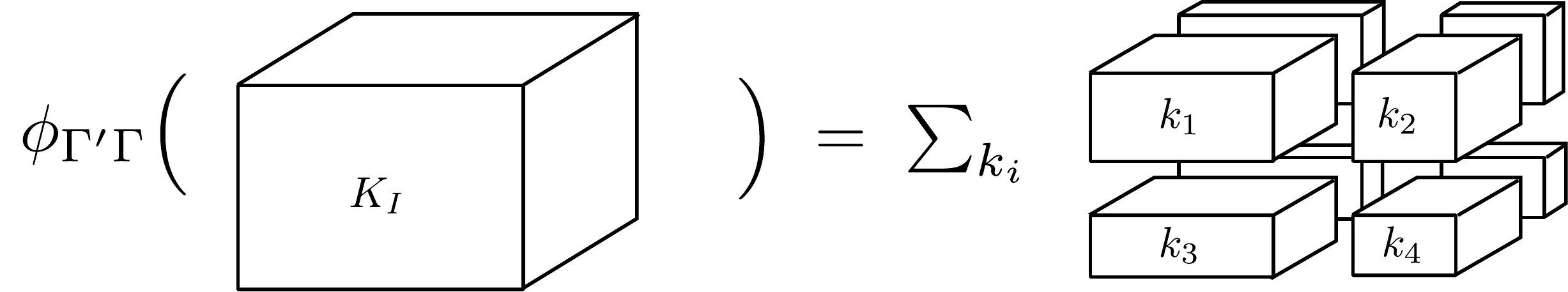}
\caption{The kinematical embedding map relates a coarse quantum cuboid to a superposition of fine ones.}\label{Fig:06}
\end{center}
\end{figure}

\noindent where $N_{\vec{K}}$ is a normalisation constant, since the embedding map is an isometry. This kinematical embedding map is an intermediate step in constructing the continuum theory.

\subsection{The physical inner product}\label{Sec:PIP}

The spin foam state sum (\ref{Eq:SpinFoamStateSum}) is used as a proposal for the physical inner product of quantum gravity. As it is defined, it provides a linear map on the boundary Hilbert space $\mathcal{H}_\Gamma$. If the boundary graph consists of two separate components $\Gamma={\Gamma}_\text{in}\cup\overline{\Gamma}_\text{out}$, then this can be rewritten as a transition map
\begin{eqnarray}\label{Eq:SpinFoamTransitionMap}
\mathcal{Z}_\Delta\;:\;\mathcal{H}_{\Gamma_\text{in}}\;\longrightarrow\;\mathcal{H}_{\Gamma_\text{out}}.
\end{eqnarray}

\noindent Here by $\overline{\Gamma}_\text{out}$ we denote the graph $\Gamma_\text{out}$ with reversed link orientations, which turns the Hilbert space to its dual. We therefore have
\begin{eqnarray}\label{Eq:PhysicalInnerProduct}
\langle\psi_\text{in}\,|\,\psi_\text{out}\rangle_\text{phys}
\;=\;
\mathcal{Z}_\Delta\left[
\psi_\text{in}\otimes\psi_\text{out}^\dag
\right].
\end{eqnarray}

\noindent The definition (\ref{Eq:PhysicalInnerProduct}) depends on the choice of $\Delta$, which needs to be removed to make the physical inner product well-defined. The traditional idea is to sum over all possible $\Delta$, which naively is not well-defined and comes with many problems \cite{Zipfel:2015diss}, while a reorganisation in terms of a GFT might be a possibility (see e.g.~\cite{Oriti:2007qd}). Another way to remove the dependence on $\Delta$ is to make the model $\Delta$-dependent to ensure mutual consistency of the physical inner products. This line of thinking led to the programme of background-independent renormalisation \cite{Oeckl:2002ia}\cite{Bahr:2014qza}\cite{Dittrich:2014ala}, and it is a way to at the same time encompass the continuum limit of the boundary Hilbert space, by refining boundary and bulk simultaneously. 

The physical inner product functions as a bona fide projector from the kinematical\footnote{The kinematical Hilbert space here is taken to be the direct sum of $\mathcal{H}_\Gamma$ for all graphs.} to the physical Hilbert space via the \emph{rigging map} $\eta:\mathcal{D}\to\mathcal{D}^*$ from a dense subspace $\mathcal{D}\subset\mathcal{H}_\text{kin}$ to its algebraic dual, given by
\begin{eqnarray}
\eta(\psi)[\phi]\;=\;\langle\psi\,|\,\phi\rangle_\text{phys}.
\end{eqnarray}

\noindent The physical Hilbert space is then derived by dividing $\mathcal{D}^*$ by the kernel of (\ref{Eq:PhysicalInnerProduct}) and completion \cite{Giulini:1998kf}. As such, physical states arise as (equivalence class of) linear combination of kinematical states on different graphs. The coefficients are given by the physical inner product itself.

In the hypercuboidal model, due to the high amount of symmetry, there are no transitions between Hilbert spaces on different graphs, i.e.~by construction a priori $\Gamma_\text{in}=\Gamma_\text{out}$. However, one can use the kinematical embedding map (\ref{Eq:EmbeddingMapQuboid}) to define a transition between differently refined graphs. As such, one can compute the physical state $\eta(\psi_\text{in})$ of a single quantum cuboid (i.e.~a graph with one node and toroidally compactified links, desctibing a torus geometry) by first embedding it into a finer graph $\Gamma'$ with more nodes, and then mapping it with the spin foam state sum. This will give the projection of $\eta(\psi_\text{in})$ to one specific graph $\Gamma'$
\begin{eqnarray}\label{Eq:PhysicalStatesProjectionToGraphs}
\eta(\psi_\text{in})_{\big|\mathcal{H}_{\Gamma'}}\;=\;
\mathcal{Z}_\Delta\circ \iota_{\Gamma'\Gamma}[\psi_\text{in}]
\end{eqnarray}

\noindent where $\mathcal{Z}_\Delta$ is interpreted as the map (\ref{Eq:SpinFoamTransitionMap}).

It is these (projections of) physical states on refined graphs $\Gamma'$ which we are considering in what follows. We are in particular interested in the entanglement entropy of these states regarding a separation of the fine graph into two halves, each containing half the nodes of $\Gamma'$.


\section{Entanglement entropy}\label{Sec:EE}

Entanglement entropy is a property of quantum states which ha received increased interest in recent years. Measuring the entanglement of degrees of freedom within a region $A$ with those outside of $A$, in particular its scaling property with increasing the size of $A$ is important. While for generic states the entanglement entropy $S_A$ grows with the volume of $A$, there is a specific class of states for which it only grows with its surface area. These arise e.g.~as the ground states of physically interesting Hamiltonians. These also play a crucial role in the renormalisation procedure for discrete systems \cite{Vidal:2007hda}.

Assume that a Hilbert space $\mathcal{H}$ can be decomposed as $\mathcal{H}=\mathcal{H}_A\otimes \mathcal{H}_B$, where $\mathcal{H}_A$ contains all degrees of freedom associated to a region $A$, and $\mathcal{H}_B$ those outside of $A$. The reduced density matrix of a state $|\psi\rangle$ w.r.t.~$A$ is then given by
\begin{eqnarray}
\hat{\rho}_A\;:=\;\text{tr}_A\big(|\psi\rangle\langle\psi|\big),
\end{eqnarray}

\noindent and the entanglement entropy $S_A$ between $A$ and $B$ is
\begin{eqnarray}
S_A\;=\;-\text{tr}_B\big(\hat{\rho}_A\,\ln\hat{\rho}_A\big).
\end{eqnarray}

\noindent If $\mathcal{H}$ does not factorise according to the region $A$ and its complement $B$, but rather takes on the form of a sum
\begin{eqnarray}\label{Eq:HS_Decomposition}
\mathcal{H}\;=\;\bigoplus_i\left(\mathcal{H}_A^{(i)}\otimes\mathcal{H}_B^{(i)}\right),
\end{eqnarray}

\noindent then 
\begin{eqnarray}
|\psi\rangle
\;=\;
\sum_i q_i |\psi_i\rangle 
\end{eqnarray}

\noindent with normalised states $|\psi_i\rangle$, each of which has an associated entanglement entropy
\begin{eqnarray}
S_A^{(i)}
\;=\;\text{tr}_{B,i}\big(\hat{\rho}_{A_i}\,\ln\hat{\rho}_{A,i}\big)
\end{eqnarray}

\noindent with $\hat{\rho}_{A,i}=\text{tr}_{A,i}\big(|\psi_i\rangle\langle\psi_i|\big)$. The total entanglement entropy associated to $A$ can  then be defined as \cite{Bianchi:2019stn}
\begin{eqnarray}\label{Eq:EE_Generalised}
S_A
\;=\;
\sum_ip_iS_A^{(i)}\;-\;\sum_ip_i\ln\big(p_i\big),
\end{eqnarray}

\noindent with $p_i=|q_i|^2$, i.e.~the weighted sum of the individual entanglement entropies, plus the von Neumann entropy of the state decomposition according to (\ref{Eq:HS_Decomposition}). It can be shown that the expression is symmetric under exchange of $A$ and $B$. 

\subsection{Entanglement entropy in LQG}

In LQG, entanglement entropy has been considered for quite some time \cite{Livine:2005mw}\cite{Livine:2006xk} \cite{Donnelly:2008vx}\cite{Delcamp:2016eya}\cite{Donnelly:2016auv}\cite{Livine:2017fgq}\cite{Feller:2017jqx}, as a spin network's degrees of freedom are localise on a graph $\Gamma$, which can be naturally split into regions. Initial computations have identified some entanglement between gauge degrees of freedom, while on the gauge-invariant level, the states carry no entanglement entropy, which can also be seen as follows: Consider the boundary Hilbert space
\begin{eqnarray}
\mathcal{H}_\Gamma\;=\;\bigoplus_{\{j_\ell\}}\left(\bigotimes_n\mathcal{H}_n^{\{j_\ell\}}\right),
\end{eqnarray}

\noindent where $\mathcal{H}_n^{\{j_\ell\}}$ is the space of intertwiners for fixed spins on the node $n$. Consider a separation of nodes of $\Gamma$ into $A$ and $B$, then any state with fixed spins 
\begin{eqnarray}
\psi\;=\;\bigotimes_n\iota_n\;=\;\bigotimes_{n\in A}\iota_n\otimes\bigotimes_{n\in B}\iota_n
\end{eqnarray}

\noindent factorises over the nodes, and has therefore vanishing entanglement entropy (\ref{Eq:EE_Generalised}). 

For linear combinations, however, the situation changes. In particular, in the hypercuboidal truncation model presented in section \ref{Sec:EPRL-Model}, the projections of physical states (\ref{Eq:PhysicalStatesProjectionToGraphs}) can be represented as finite linear combinations of spin networks. We will therefore write them as states within the kinematical Hilbert spaces. For a separation of nodes into regions $A$ and $B=N(\Gamma)\backslash A$, the entanglement entropy  $S_A$ will in general not vanish. 

We consider three cases in what follows:

\subsection{Case 1}

First we consider the projection of one quantum cuboid state to two quantum cuboids: $\Gamma_\text{in}$ consists of one node, with three loops as links, defining a toroidally compactified geometry depending on thee spins $K_1$, $K_2$, and $K_3$. The refined state $\Gamma_\text{out}$ consists of two nodes connected by one link, resulting from one cuboid dissected into two. There are five independent spins $k_1,\ldots, k_5$. The (projected) physical state is, after normalisation, given by
\begin{eqnarray}\label{Eq:Case_1_State}
\eta[\iota_{K_1,K_2,K_3}]\;=\;\frac{1}{N_{K_1,K_2,K_3}^{(\alpha)}}\sum_{k_1,\ldots,k_5}c_{k_1,\ldots,k_5}^{(\alpha)}\iota_{k_1,k_2,k_3}\otimes\iota_{k_1,k_4,k_5}
\end{eqnarray}

\begin{figure}
\begin{center}
\includegraphics[scale=0.45]{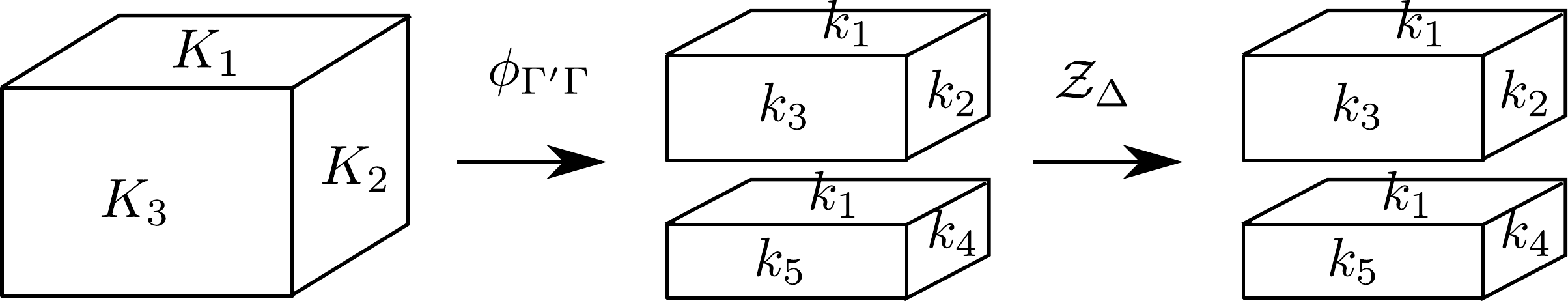}
\caption{In case 1, one quantum cuboid is transitioned into two.}\label{Fig:07}
\end{center}
\end{figure}

\noindent with
\begin{eqnarray}\label{Eq:Case_1_Coefficient}
c_{k_1,\ldots,k_5}^{(\alpha)}
\;=\;
\sum_{k_6,\ldots,j_{11}}\hat{\mathcal{A}}_{k_1,k_2,k_3,k_6,k_7,k_8}^{(\alpha)}\hat{\mathcal{A}}_{k_1,k_4,k_5,k_9,k_{10},k_{11}}^{(\alpha)}
\end{eqnarray}

\begin{figure}
\begin{center}
\includegraphics[scale=0.45]{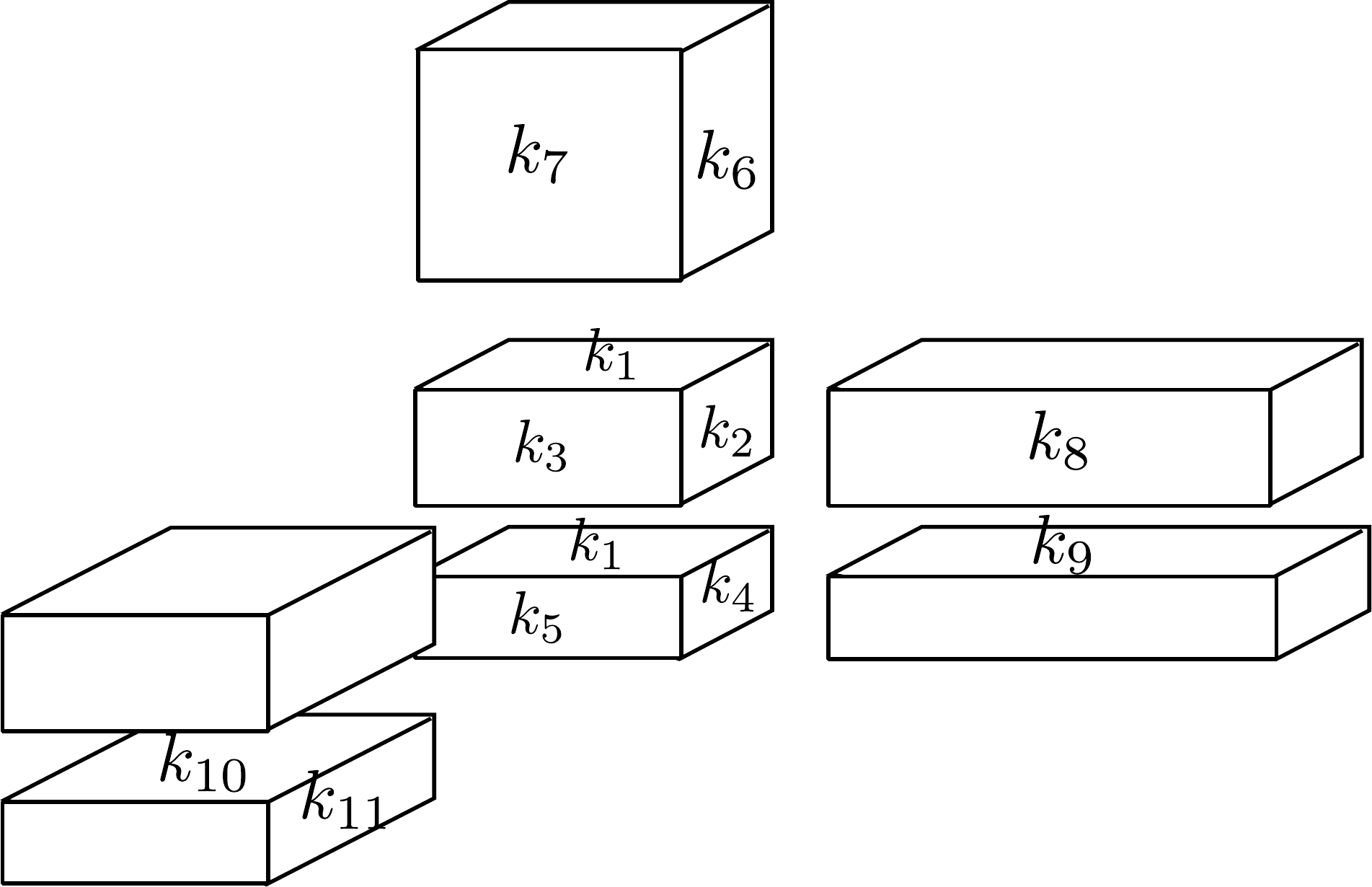}
\caption{The bulk of case 2 consists of two hypercuboids, whose boundary is partially depicted. There are a priori $k_1,\ldots ,k_{11}$ as possible spins.}\label{Fig:08}
\end{center}
\end{figure}

\noindent and 
\begin{eqnarray}\label{Eq:Case_1_Normalisation}
\left(N_{K_1,K_2,K_3}^{(\alpha)}\right)^2
\;=\;
\sum_{k_1,\ldots,k_5}
\left|c_{k_1,\ldots,k_5}^{(\alpha)}\right|^2.
\end{eqnarray}

\noindent From the properties of the kinematical embedding map (\ref{Eq:EmbeddingMapQuboid}) we can see that the sums in (\ref{Eq:Case_1_State}) and (\ref{Eq:Case_1_Normalisation}) is restricted to
\begin{eqnarray}
K_1\;=\;k_1,\qquad K_2\;=\;k_2+k_4,\qquad K_3\;=\;k_3+k_5.
\end{eqnarray}

\noindent Due to correct glueing of the 4-dim hypercuboids, the sum (\ref{Eq:Case_1_Coefficient}) has to range over
\begin{eqnarray}
k_6\;=\;k_9,\qquad k_7\;=\;k_{10}
\end{eqnarray}

\noindent We consider two more simplifications:
\begin{itemize}
\item \textbf{Volume-simplicity:} We impose the Hopf-link-volume-simplicity constraint discussed in section \ref{Sec:EPRL-Model} \cite{Bahr:2017ajs}, leading to the additional conditions
\begin{equation}
\begin{aligned}\label{Eq:Case_1_Volume_Simplicity_Constraints}
k_1k_8\;&=&\;k_2k_7\;=\;k_3k_6,\\[5pt]
k_1k_{11}\;&=&\;k_4k_{10}\;=\;k_5k_9.
\end{aligned}
\end{equation}

\noindent Note that the Hopf-link constraints impose face-matching for each vertex separately, but this also leads to face-matching on the boundary Hilbert spaces, effectively restricting the Hilbert spaces to those which correspond to torsion-free geometries \cite{Bahr:2015gxa}. 

\item \textbf{Ischoric transition:} We furthermore restrict the allowed transitions to those which fix the total $4d$-volume $V$, which in this case is given by
\begin{eqnarray}
V\;=\;\frac{k_1k_8+k_2k_7+k_3k_6}{3}\,+\,\frac{k_1k_{11}+k_4k_{10}+k_5k_9}{3}
\end{eqnarray} 
\noindent Note that this leads to the constraints 
\begin{eqnarray}
k_8\,+\,k_{11}\;=\;K_6, \qquad k_6\;=\;K_4,\qquad k_7\;=\;K_5
\end{eqnarray}
\noindent where $K_4,K_5,K_6$ are the spins of the coarse hypercuboid, which satisfy
\begin{eqnarray}
V\;=\;K_1K_6\;=\;K_2K_5\;=\;K_3K_4
\end{eqnarray}
\noindent due to the volume-simplicity constraint.
\end{itemize}
\noindent Together with (\ref{Eq:Case_1_Volume_Simplicity_Constraints}), this leads in total to a sum over one single boundary spin and no bulk spin, i.e.~
\begin{eqnarray}\label{Eq:Case_1_State_Simplified}
\eta[\iota_{K_1,K_2,K_3}]
\;=\;\frac{1}{N^{(\alpha)}}\sum_{k=0}^{K_2}c_k^{(\alpha)}\;\iota_{K_1,k,kK_3/K_2}\otimes\iota_{K_1,(K_2-k),(K_2-k)K_3/K_2}
\end{eqnarray}

\noindent with
\begin{eqnarray}\label{Eq:Case_1_Coefficient_Simplified}
c_k^{(\alpha)}\;=\;
\hat{\mathcal{A}}^{(\alpha)}_{K_1,k,k \frac{K_3}{K_2},\frac{V}{K_3},\frac{V}{K_2},k\frac{V}{K_1K_2}}
\hat{\mathcal{A}}^{(\alpha)}_{K_1,K_2-k,(K_2-k) \frac{K_3}{K_2},\frac{V}{K_3},\frac{V}{K_2},(K_2-k)\frac{V}{K_1K_2}}
\end{eqnarray}

\noindent and
\begin{eqnarray}\label{Eq:Case_1_Normalisation_Simplified}
\left(N^{(\alpha)}\right)^2\;=\;\sum_k \left| c_k^{(\alpha)} \right| ^2.
\end{eqnarray}

\noindent From the form (\ref{Eq:Case_1_State_Simplified}) one can see that the physical state lies in the direct product Hilbert space for fixed $k_1=K_1$:
\begin{eqnarray}
\eta[\iota_{K_1,K_2,K_3}]\;\in\;
\mathcal{H}_A\otimes\mathcal{H}_B
\end{eqnarray}
\noindent with
\begin{eqnarray}
\mathcal{H}_A\;=\;\bigotimes_{k_2,k_3}\text{span}\big(\iota_{k_1,k_2,k_3}\big),\qquad
\mathcal{H}_B\;=\;\bigotimes_{k_4,k_5}\text{span}\big(\iota_{k_1,k_4,k_5}\big),\qquad
\end{eqnarray}

\noindent With this, using (\ref{Eq:Case_1_State_Simplified}) and tracing subsequently over degrees of freedom in $A$ and $B$, one straightforwardly arrives at

\begin{eqnarray}\label{Eq:Case_1_EE}
S_A^{(\alpha)}\;=\;\text{ln}\left(N^{(\alpha)}\right)^2
\;-\;
\frac{1}{\left(N^{(\alpha)}\right)^2}\sum_k\;\left| c_k^{(\alpha)} \right| ^2\text{ln}\left(\left| c_k^{(\alpha)} \right| ^2\right).
\end{eqnarray}

\subsection{Case 2}

\begin{figure}
\begin{center}
\includegraphics[scale=0.45]{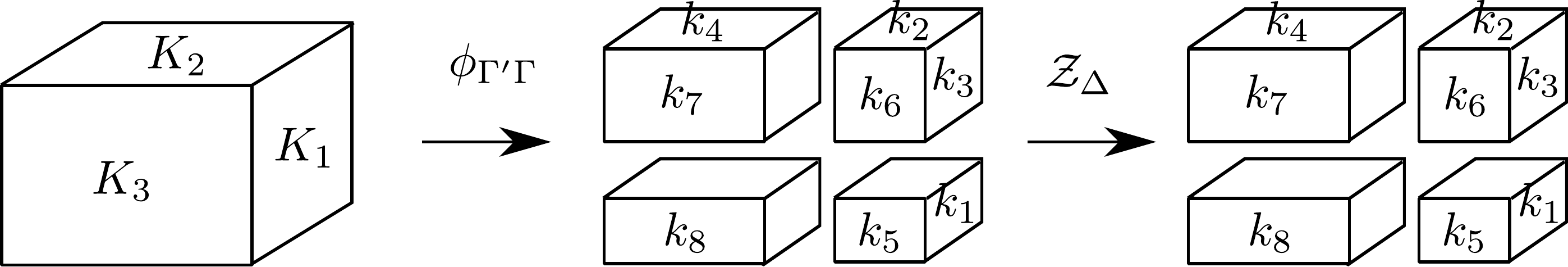}
\caption{In case 2, one quantum cuboid is transitioned into four, two of which form the regions $A$ and $B$, respectively. In this figure these are the two right and the two left ones.}\label{Fig:09}
\end{center}
\end{figure}

In the second case we consider the subdivision of one quantum cuboid into four. Two of them, respectively, form the regions $A$ and $B$ (see figure \ref{Fig:09}). The final state therefore is of the form
\begin{eqnarray}\label{Eq:Case_2_PhysicalStateSimple}
\eta[\psi_{K_1,K_2,K_3}]
\;=\;
\sum_{k_i}\,c_{k_i}^{(\alpha)}\iota_{k_1,k_2,k_5}\otimes\iota_{k_3,k_2,k_6}\otimes\iota_{k_1,k_4,k_8}
\otimes\iota_{k_3,k_4,k_7}
\end{eqnarray}

\noindent Due to the embedding map (\ref{Eq:EmbeddingMapQuboid}) the coefficients will be zero unless
\begin{eqnarray}\label{Eq:Case_2_EmbeddingMapConditions}
\begin{aligned}
k_1+k_3\;&=&\;K_1,\\[5pt]
k_2+k_4\;&=&\;K_2,\\[5pt]
k_5+k_6+k_7+k_8\;&=&\;K_3,
\end{aligned}
\end{eqnarray}

\noindent while geometricity will enforce 
\begin{eqnarray}\label{Eq:Case_2_Geometricity_Conditions}
\begin{aligned}
k_5\;=\;\frac{k_1k_2}{K_1K_2}K_3,\qquad k_6\;=\;\frac{k_3k_2}{K_1K_2}K_3,\\[5pt]
k_7\;=\;\frac{k_3k_4}{K_1K_2}K_3,\qquad k_8\;=\;\frac{k_1k_4}{K_1K_2}K_3.
\end{aligned}
\end{eqnarray}

\noindent From this one can see that $k_1$ and $k_2$ are the only independent variables in the coefficients of the physical state (\ref{Eq:Case_2_PhysicalStateSimple}). 

The bulk consists of four hypercuboids, i.e. in the 2-complex there are 4 vertices. Similarly to case 1, the isochoric constraint of fixing the total $4$-volume $V=V_1+V_2+V_3+V_4$, which results in
\begin{eqnarray*}
c_{k_i}^{(\alpha)}\;\equiv\;c_{k_1,k_2}^{(\alpha)}\;&=&\;
\hat{\mathcal{A}}_{k_1,k_2,k_5,K_4,\frac{k_1}{K_1}K_5, \frac{k_2}{K_2}K_6}
\hat{\mathcal{A}}_{k_3,k_2,k_6,K_4,\frac{K_1-k_1}{K_1}K_5, \frac{k_2}{K_2}K_6}\\[5pt]
&&\quad \times 
\hat{\mathcal{A}}_{k_1,k_4,k_8,K_4,\frac{k_1}{K_1}K_5,  \frac{K_2-k_2}{K_2}K_6}
\hat{\mathcal{A}}_{k_3,k_4,k_7,K_4,\frac{K_1-k_1}{K_1}K_5,  \frac{K_2-k_2}{K_2}K_6}
\end{eqnarray*}

\noindent with
\begin{eqnarray}
K_4\;=\;\frac{V}{K_3},\quad
K_5\;=\;\frac{V}{K_2},\quad
K_6\;=\;\frac{V}{K_1}.
\end{eqnarray}

\noindent With (\ref{Eq:Case_2_EmbeddingMapConditions}) and (\ref{Eq:Case_2_Geometricity_Conditions}) these can be written entirely in terms of $k_1$ and $k_2$. In the graph $\Gamma_\text{out}$ there are four nodes, two of which we regard as being in region $A$, while the other two are in region $B$ (see figure \ref{Fig:09}). Therefore the physical state is in
\begin{eqnarray}
\eta[\psi_{K_1,K_2,K_3}]
\;\in\;
\sum_{k_1}\mathcal{H}_{k_1}^{(A)}\otimes\mathcal{H}_{k_1}^{(B)}
\end{eqnarray}
\noindent with
\begin{eqnarray}
\mathcal{H}_{k_1}^{(A)}
\;&=&\;\bigoplus_{k_2,k_5,k_6}\text{span}\left(
\iota_{k_1,k_2,k_5}\otimes \iota_{K_1-k_1,k_2,k_6}
\right)
\\[5pt]
\mathcal{H}_{k_1}^{(B)}
\;&=&\;
\bigoplus_{k_4,k_7,k_8}\text{span}\left(
\iota_{k_1,k_4,k_8}\otimes \iota_{K_1-k_1,k_4,k_7}
\right)
\end{eqnarray}

\noindent The physical state is therefore not in a Hilbert space which is a tensor product over the regions $A$ and $B$, but rather a direct sum of those products. We therefore use the generalised expression (\ref{Eq:EE_Generalised} for the entanglement entropy. We write
\begin{eqnarray}
\eta[\psi_{K_1,K_2,K_3}]
\;=\;\sum_{k_1}q_{k_1}\psi_{k_1}
\end{eqnarray}

\noindent with real coefficients
\begin{eqnarray}
q_{k_1}
\;=\;\frac{N_{k_1}}{N}
\end{eqnarray}

\noindent and normalised states
\begin{eqnarray}
\psi_{k_1}
\;=\;
\frac{1}{N_{k_1}}\sum_{k_2}c_{k_1,k_2}^{(\alpha)}\iota_{k_1,k_2,k_5}\otimes\iota_{k_3,k_2,k_6}\otimes\iota_{k_1,k_4,k_8}
\otimes\iota_{k_3,k_4,k_7},
\end{eqnarray}

\noindent where $k_3,\ldots, k_8$ are determined by $k_1,k_2$ via (\ref{Eq:Case_2_EmbeddingMapConditions}) and (\ref{Eq:Case_2_Geometricity_Conditions}), and where
\begin{eqnarray}
\begin{aligned}
N_{k_1}^2
\;&=&\;\sum_{k_2}\left|c_{k_1,k_2}^{(\alpha)}\right|^2\\[5pt]
N^2\;&=&\;\sum_{k_1,k_2}\left|c_{k_1,k_2}^{(\alpha)}\right|^2
\end{aligned}
\end{eqnarray}

\noindent For a fixed $k_1$, the entanglement entropy $S_{A}^{(k_1)}$ can be computed similarly to (\ref{Eq:Case_1_EE}), and yields
\begin{eqnarray}
S_{A}^{(k_1)}
\;=\;
\text{ln}\left(N_{k_1}^2\right)\,-\,\frac{1}{N_{k_1}^2}\sum_{k_2}\left|c_{k_1,k_2}^{(\alpha)}\right|^2\text{ln}\left|c_{k_1,k_2}^{(\alpha)}\right|^2
\end{eqnarray}

\noindent The total entanglement entropy, with (\ref{Eq:EE_Generalised}), can be computed to be
\begin{eqnarray}\label{Eq:Case_2_EE}
\begin{aligned}
S_A\;=&\;\sum_{k_1}\frac{N_{k_1}^2}{N^2}\left[
\text{ln}\,N_{k_1}^2\,-\,\frac{1}{N_{k_1}^2}\sum_{k_2}\left|c_{k_1,k_2}^{(\alpha)}\right|^2\text{ln}\left|c_{k_1,k_2}^{(\alpha)}\right|^2
\right]\,-\,\sum_{k_1}\frac{N_{k_1}^2}{N^2}\text{ln}\frac{N_{k_1}^2}{N^2}\\[5pt]
=&\;\text{ln}\,N^2\,-\,\frac{1}{N^2}\sum_{k_1,k_2}\left|c_{k_1,k_2}^{(\alpha)}\right|^2\text{ln}\left|c_{k_1,k_2}^{(\alpha)}\right|^2.
\end{aligned}
\end{eqnarray}

\section{Numerical computations}\label{Sec:Numerics}

In what follows we will present numerical results on the entanglement entropy for cases 1 and 2 from the last section. We work entirely in the large-spin-.asymptotic regime, and assume that we can neglect the boundary contributions from small spins\footnote{In particular in the isochoric case this has proven a good assumption, see \cite{Bahr:2015gxa} for a discussion.}. 

We assume that the spins are so large that the summations can be turned into integrals, even when taking the EPRL-FK quantisation condition (\ref{Eq:SimplicitySpinCondition}) into account\footnote{See also discussion in section \ref{Sec:EPRL-Model}.}, and one can use the asymptotic expressions for the amplitudes  (\ref{Eq:VerteAmplitude} -- \ref{Eq:FaceAmplitude}). Since the asymptotic formulas are homogenous under simultaneous scaling of the spins
\begin{eqnarray}
\hat{A}^{(\alpha)}(\lambda k_1,\ldots, \lambda k_6)\;=\;\lambda^\beta
\hat{A}^{(\alpha)}(k_1,\ldots, k_6)
\end{eqnarray} 

\noindent with $\beta = 12\alpha-9$ \cite{Bahr:2015gxa}, the entanglement entropy $S_A$ has a specific scaling behaviour with respect to $K_1, K_2,K_3$. 

First we treat case 1: We have
\begin{eqnarray}
S_A^{(\alpha),\lambda}
\;=\;\text{ln} N^2\;-\;\frac{1}{N^2}\int_0^{K_1} dk \;\left |c_{k}^{(\alpha)}\right|^2\;\text{ln}\left |c_{k}^{(\alpha)}\right|^2
\end{eqnarray}
\noindent with 
\begin{eqnarray}
N^2\;=\;\int_0^{K_1}dk\;\left |c_{k}^{(\alpha)}\right|^2.
\end{eqnarray}

\noindent Denoting by $S_A^{(\alpha),\lambda}$ the entanglement entropy of the physical state $\eta[\iota_{\lambda K_1,\lambda K_2,\lambda K_3}]$, we get
\begin{eqnarray}\label{Eq:EE_Scaling_01}
S_A^{(\alpha),\lambda}\;=\;S_A^{(\alpha)}\;+\;\text{ln}\lambda\qquad \text{(Case 1)}
\end{eqnarray}

\noindent Similarly, the scaling of the entanglement entropy in case 2 can be computed as
\begin{eqnarray}\label{Eq:EE_Scaling_02}
S_A^{(\alpha),\lambda}
\;=\;S_A^{(\alpha)}\;+\;\text{ln}\lambda^2\qquad \text{(Case 2)}.
\end{eqnarray}

\noindent This scaling\footnote{It should be noted at this point that this scaling is in no way related to the question of whether the entanglement entropy scales with the area or the volume. In particular, even though the spins $K_i$ are related to the areas, equation (\ref{Eq:EE_Scaling_01}) should not be interpreted as an area scaling law. Such a law could only be inferred by keeping the spins $K_i$ fixed and increasing the lattice sites, and computing the scaling of $S_A$ with regards to this increase. This is an interesting question outside the scope of this article,  which we leave for future investigations.} of $S_A$ is very useful when it comes to numerical investigations.

\begin{figure}
\begin{center}
\includegraphics[scale=0.75]{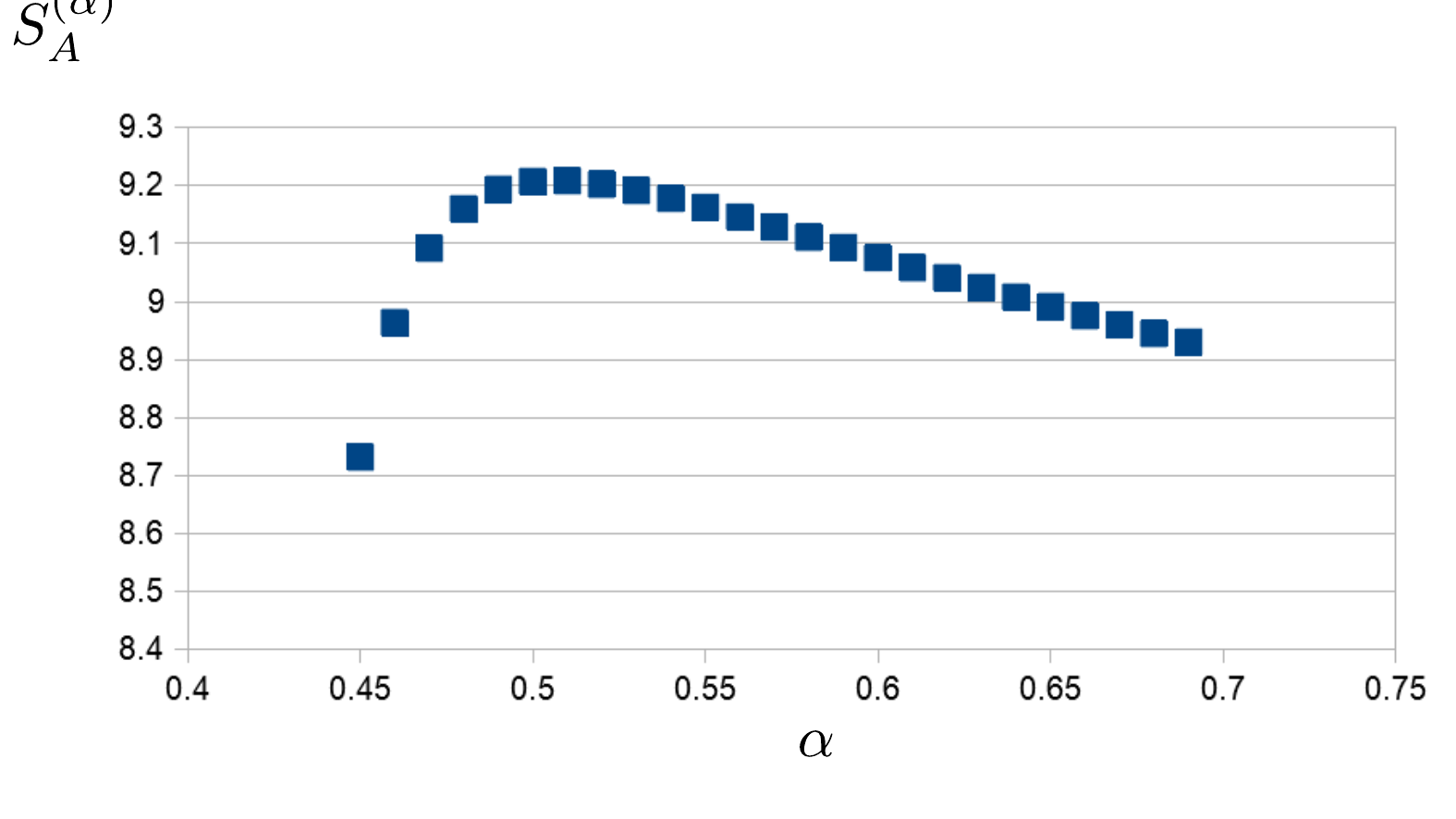}
\caption{Entanglement entropy $S^{(\alpha)}$ depending on $\alpha$, for case 1 with $K_1=K_2=K_3=10^5$, and $V=10^{10}$.}\label{Fig:10}
\end{center}
\end{figure}

\begin{figure}
\begin{center}
\includegraphics[scale=0.75]{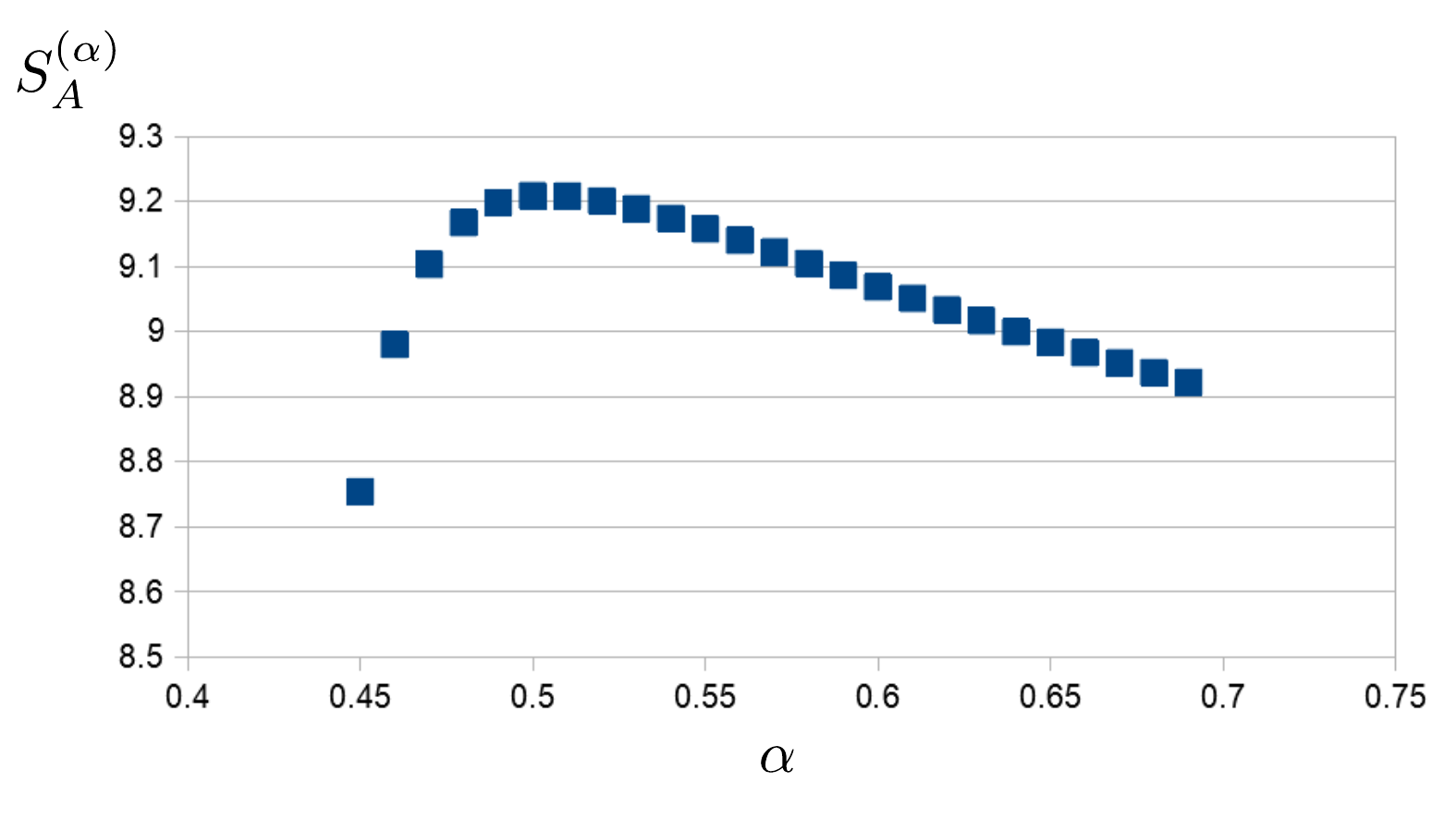}
\caption{Entanglement entropy $S^{(\alpha)}$ depending on $\alpha$, for case 1 with $K_1=K_2=K_3=10^5$, and $V=2\cdot 10^{10}$.}\label{Fig:11}
\end{center}
\end{figure}

\begin{figure}
\begin{center}
\includegraphics[scale=0.75]{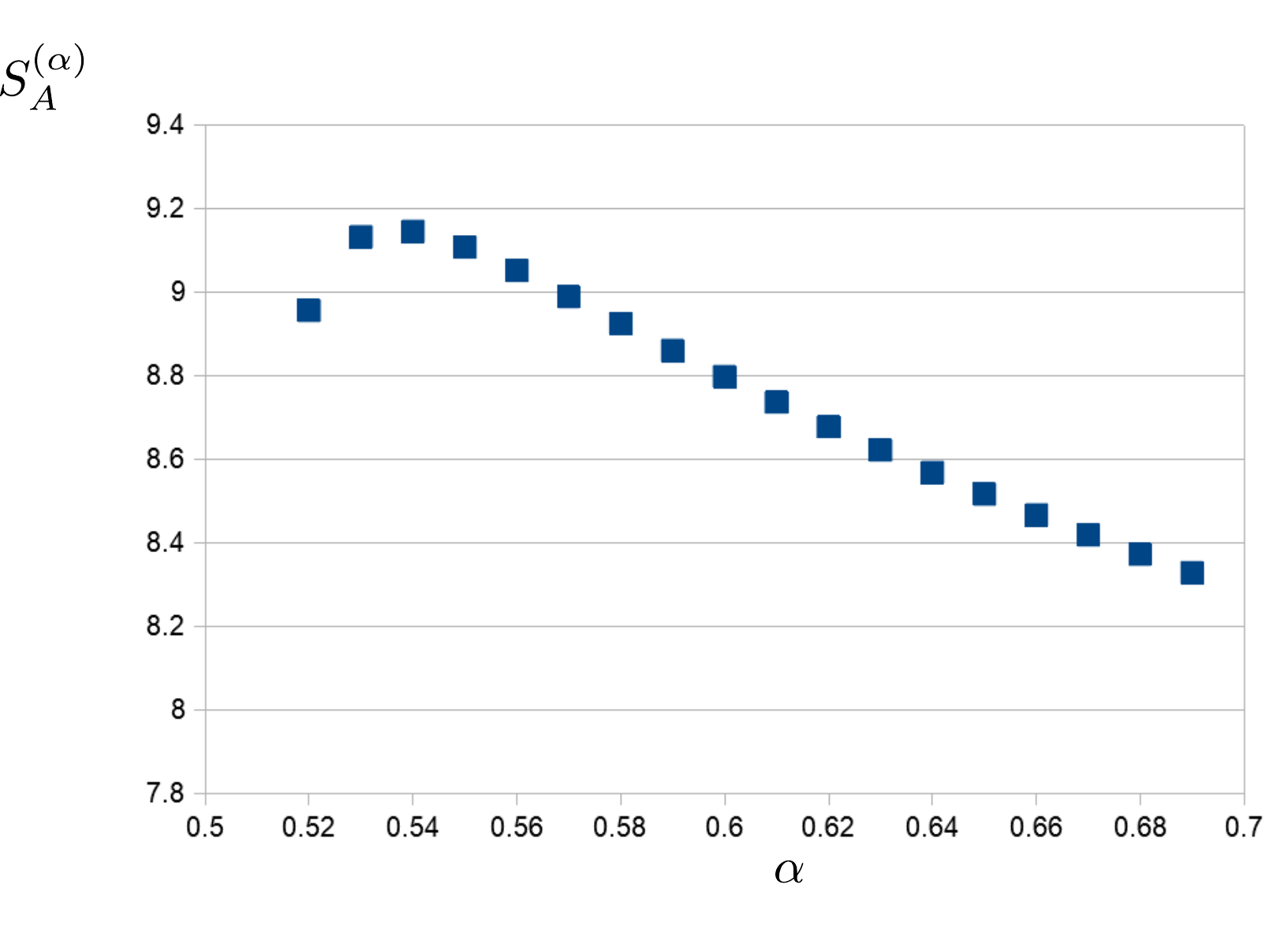}
\caption{Entanglement entropy $S^{(\alpha)}$ depending on $\alpha$, for case 2 with $K_1=K_2=K_3=10^5$, and $V= 10^{10}$.}\label{Fig:12}
\end{center}
\end{figure}

Using the scaling behaviour (\ref{Eq:EE_Scaling_01} -- \ref{Eq:EE_Scaling_02}), we can numerically evaluate the integrals (\ref{Eq:Case_1_EE}), (\ref{Eq:Case_2_EE}) for different values of initial spins $K_1$, $K_2$, $K_3$. We use numerical integration techniques from the GNU scientific library (GSL), which can be straightforwardly implemented in $C++$. In figures \ref{Fig:10} -- \ref{Fig:12}, we present the dependence on the coupling constant $\alpha$ for various fixed $K_i$, for either case.

One can see a clear behaviour of the entanglement entropy, which has a maximum around 
\begin{eqnarray}
\alpha_\text{max}\;\approx\; 0.5
\end{eqnarray}
\noindent where the precise value depends on the (ratios of the) $K_i$. The contour is qualitatively robust, however, and $S_A$ decreases rapidly for smaller values of $\alpha$. It also decreases for larger values, albeit more slowly.

\begin{figure}
\begin{center}
\includegraphics[scale=0.55]{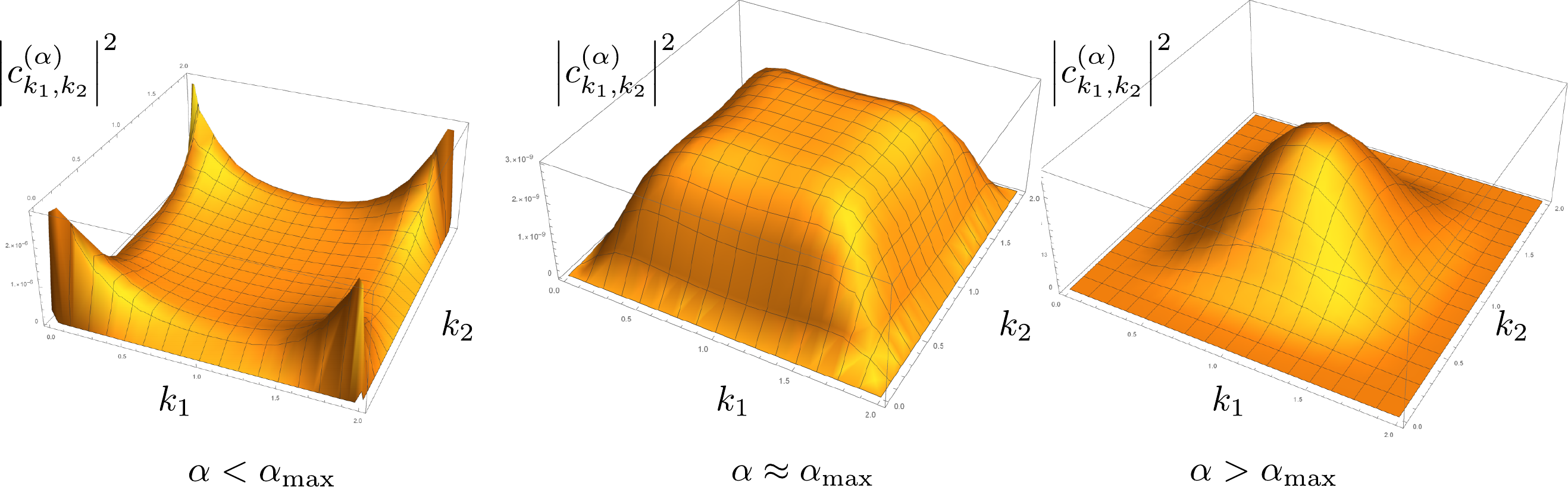}
\caption{Coefficient $\left|c_{k_1,k_2}^{(\alpha)}\right|^2$  of the physical state in case 2, for various $\alpha$.}\label{Fig:13}
\end{center}
\end{figure}

This behaviour can be directly understood when taking a closer look at the coefficients $c_k^{(\alpha)}$ and $c_{k_1,k_2}^{(\alpha)}$. In figure \ref{Fig:13} we have depicted the graph of the coefficients for different $\alpha$. One can clearly see that for small $\alpha$, $c_{k_1,k_2}^{(\alpha)}$ is sharply peaked around few values of $k_1,k_2$, indicating that the physical state is a superposition of only few states with definite spins, resulting in low entanglement entropy. For large values of $\alpha$, the coefficients are also concentrated around specific values of spins $k_i$, albeit with a larger spread. There is an intermediate regime in which the coefficients are spread out over a much larger regions of spins, indicating that the physical state is a coherent superposition of a large number of different spins $k_1, k_2$, which leads to a large entanglement entropy. 

The geometric interpretation of this becomes clear when one considers the 3-volume. Since we are working in the isochoric framework, the total 4-volume $V$ is fixed in the transition between the in- and out-state, and hence, in the hypercubic setting, also the spatial 3-volume is. Different values of $k_i$ therefore correspond to different states in which the 3-volume
\begin{eqnarray}
V^{(3)}
\;=\;\sqrt{K_1 K_2 K_3}
\end{eqnarray}
\noindent  is distributed differently between the regions $A$ and $B$. In the case 1 one can directly see that
\begin{eqnarray}
V_A
\;&=&\;\frac{k}{K_1}V^{(3)} \\[5pt]
V_B
\;&=&\;\frac{K_1-k}{K_1}V^{(3)}
\end{eqnarray}

\noindent and for case 2:
\begin{eqnarray}
V_A
\;&=&\;
\sqrt{k_1k_2k_5}+\sqrt{k_2k_3k_6}\;=\;\frac{k_2}{K_2}V^{(3)}
 \\[5pt]
V_B
\;&=&\;\sqrt{k_1k_4k_8}+\sqrt{k_3k_4k_7}\;=\;\frac{K_2-k_2}{K_2}V^{(3)}
\end{eqnarray}

\noindent From this and figure \ref{Fig:13}, one can see that the main contribution to the physical state for $\alpha < \alpha_\text{max}$ comes from either of the four extremal cases
\begin{eqnarray}
(k_1,k_2)\;=\;(0,0),\;(0, K_2),\;(K_1,0),\;(K_1,K_2).
\end{eqnarray}

\noindent Two of these cases correspond to $V_A=0$, $V_B=V^{(3)}$, and two to $V_A=V^{(3)}$, $V_B=0$. 

Conversely, the regime $\alpha\,>\,\alpha_\text{max}$ leads to the main contribution coming from an area around
\begin{eqnarray}
(k_1,k_2)\;\approx\;\left(\frac{K_1}{2},\frac{K_2}{2}\right),
\end{eqnarray}

\noindent which corresponds to
\begin{eqnarray}
V_A\;\approx\;V_B\;\approx\;\frac{V^{(3)}}{2},
\end{eqnarray}

\noindent i.e.~where the 3-volume is distributed equally between the regions $A$ and $B$. 

Between these two extremal cases there is an intermediate regime in which almost all possible distributions of 3-volume among $A$ and $B$ occur with roughly equal probability in the physical state. This is the state with maximal entanglement entropy. 

These features appear to occur for various different initial spins $K_i$, and both cases 1 and 2. From investigations with larger lattices it can be expected that the peak of $S_A$ at $\alpha=\alpha_\text{max}$ becomes even more pronounced as the number of lattice sites increases. This is due to the fact that the  coefficients $c^{(\alpha)}_{k_i}$ of the physical states, as product of more and more amplitudes $\hat{\mathcal{A}}$, become more and more sharply peaked.

\section{Summary and discussion}\label{Sec:Summary}

In this article we have considered the entanglement entropy of physical states in the EPRL-FK spin foam model, where we have worked in the hypercuboidal truncation. In the large-spin regime this model depends on only one parameter $\alpha$, and has a manageable set of degrees of freedom, which makes this truncation an interesting toy model for the full, untruncated, theory. 

Rather than single spin network functions, we considered physical states as given by the spin foam transition, which in this setting arise as linear combination of spin networks. The boundary graph are dual to cubic 3d lattices, describing a Cauchy surface with the topology of a 3-torus. We have considered graphs with an even number of nodes, such that the ``universe'' could be separated into two similar regions $A$ and $B$. We then numerically computed the entanglement entropy $S_A$ of these states with regards to the separation of space into $A$ and $B$. 

We were specifically interested in the dependence of $S_A$ on the parameter $\alpha$, for different physical states. We have found that $S_A$, generically has a maximum around $\alpha_\text{max}\approx 0.5$. It therefore lies in the ```critical regimes'' of $\sim 0.5 - 0.65$, where the model generically undergoes a qualitative change in behaviour. This regime has been found to have several interesting features, and in particular contains the fixed point of the background-independent RG flow \cite{Bahr:2016hwc}\cite{Bahr:2017eyi}. This is the point where the diffeomorphism symmetry gets restored, which is broken in the EPRL-FK model due to the discretisation \cite{Bahr:2009qc}\cite{Bahr:2015gxa}. 

It is this restoration of diffeomorphism symmetry which we conjecture to be the reason for the maximising of the entanglement entropy in this region. In particular, the physical states are superpositions of spin network functions which -- in the large spin regime we are considering -- are all on the same orbit of the classical vertex translation symmetry group, which is the lattice version of the diffeomorphism group which arises e.g.~in Regge Calculus \cite{Rocek:1982fr}\cite{FreidelLouapreDiffeo2002}\cite{Bahr:2009qc}\cite{Dittrich:2012qb}\cite{Bahr:2015gxa}. For $\alpha\approx\alpha_\text{max}$, these diffeomorphically equivalent degrees of freedom of the kinematical (spin-network) states are becoming maximally entangled with one another, which here arises as another feature of the restoration of diffeomorphisms at the RG  fixed point. It can in particular be regarded to be a consequence of the fact that the subdivision of space into two regions $A$ and $B$ is being performed not with regards to any physical property of the system, but with regards to the nodes of the graph, which functions as external structure here. Thus, the separation of space into $A$ and $B$ is not diffeomorphism-invariant, in line with the discussion in \cite{Bahr:2015gxa}. 

This findings could be used for future investigations, also in the full theory. Those points in parameter space with maximal entanglement entropy indicate interesting behaviour with regards to the diff symmetry, and therefore could be used to find e.g.~fixed points of the RG flow. This would be far less effort than computing the RG flow in the full theory, which is still an unsolved problem. 

In the future, it would be interesting to check whether the behaviour of the entanglement entropy is persistent when relaxing the hypercuboidal truncation of the model. In particular including curvature degrees of freedom is necessary in order to solidify the results of this analysis. We hope to come back to this point in another article.

\section*{Acknowledgements}
 This work was funded by the project BA 4966/1-2 of the German Research Foundation (DFG).

\bibliography{VolumeBib}
\bibliographystyle{ieeetr}

\end{document}